# A Comment on the calculation of the Lie algebra cocycle in the book *Loop Groups*.

Dan Solomon
Rauland-Borg Corporation
Mount Prospect, IL
Email: dan.solomon@rauland.com
March 22, 2013

**Abstract.**

An expression for the Lie algebra cocycle corresponding to the central extension of the group $GL_{res}(H)$ is derived in the book *Loop Groups* [1]. It will be shown in this paper that some of the terms in this expression are ambiguous and in order to be evaluated correctly the trace operation must be properly defined.

## 1. Introduction.

The Lie algebra cocycle corresponding to the central extension of the group $GL_{res}(H)$ is derived in Section (6.6) of the book *Loop Groups* [1]. An expression for the cocycle is given in Proposition (6.6.5) of [1]. (For the definition and further discussion of the group $GL_{res}$ see Chapter 6 of Ref. [1].) The purpose of this paper is to examine this expression for the cocycle and show that its evaluation depends on the way the trace operation is defined.

We consider a Hilbert space defined on a circle. For this case the basis vectors are given by,

$$f_n(\theta) = \frac{e^{-in\theta}}{\sqrt{2\pi}} \qquad (1.1)$$

where $0 \le \theta \le 2\pi$ and $n$ is an integer. Define the projection operators $P_+$ and $P_-$ where,

$$P_+ f_n(\theta) = \begin{cases} f_n(\theta) \text{ if } n \ge 0 \\ 0 \text{ if } n < 0 \end{cases} \quad \text{and} \quad P_- = 1 - P_+ \qquad (1.2)$$

From proposition (6.6.5) of *Loop Groups* [1] we have the following,

$$F_1 = F_2 = F_3 \tag{1.3}$$

where,

$$F_1 = trace([a_1, a_2] - a_3) \tag{1.4}$$

$$F_2 = trace(c_1 b_2 - b_1 c_2) \tag{1.5}$$

$$F_3 = \frac{1}{4} trace(J[J, A_1][J, A_2]) \tag{1.6}$$

where $J = P_+ - P_-$ and,

$$a_1 = P_+ A_1 P_+, \; b_1 = P_+ A_1 P_-, \; c_1 = P_- A_1 P_+ \tag{1.7}$$

$$a_2 = P_+ A_2 P_+, \; b_2 = P_+ A_2 P_-, \; c_2 = P_- A_2 P_+ \tag{1.8}$$

$$a_3 = P_+ [A_1, A_2] P_+ \tag{1.9}$$

where $A_1$ and $A_2$ are operators on the Hilbert space such that $[J, A_1]$ and $[J, A_2]$ are Hilbert-Schmidt.

It will be shown that for Eq. (1.3) to be true the trace operation must be properly defined. In particular we cannot assume that the cyclicity of the trace holds for the above quantities.

## 2. An inconsistency.

Using the above expressions the quantities $F_1$ and $F_2$ can be rewritten as,

$$F_1 = trace([a_1, a_2] - P_+ [A_1, A_2] P_+) \tag{2.1}$$

$$F_2 = trace(P_- A_1 P_+ A_2 P_- - P_+ A_1 P_- A_2 P_+) \tag{2.2}$$

To show that there is a possible problem with evaluating the above expressions let us calculate $F_1$ and $F_2$ for the case where,

$$A_1 = \cos(j\theta) \text{ and } A_2 = \sin(j\theta) \tag{2.3}$$

where $j$ is a positive integer. In this case $A_1$ and $A_2$ commute therefore $[A_1, A_2] = 0$ and $F_1$ becomes,

$$F_1 = trace(a_1 a_2 - a_2 a_1) \tag{2.4}$$

The cyclicity of the trace is the relationship $trace(XY) = trace(YX)$. Using this in the above we obtain,

$$F_1 = 0 \tag{2.5}$$

To evaluate $F_2$ we note that the effect of the projection operators $P_+$ and $P_-$ on some function $g(\theta)$ is,

$$P_+ g(\theta) = \sum_{n=0}^{\infty} f_n \langle f_n^* g \rangle \text{ and } P_- g(\theta) = \sum_{n=-1}^{-\infty} f_n \langle f_n^* g \rangle \tag{2.6}$$

where,

$$\langle f(\theta) \rangle = \int_0^{2\pi} f(\theta) d\theta$$

The trace of a matrix $A_{ij}$ is defined by,

$$trace(A_{ij}) = \sum_{n=-\infty}^{n=+\infty} A_{nn} \tag{2.7}$$

From the definition of the trace and the projection operator (2.2) becomes,

$$F_2 = \frac{1}{(2\pi)^2} \left\{ \begin{array}{l} \sum_{n=1}^{\infty} \sum_{m=0}^{\infty} \left( \langle e^{-in\theta} A_1 e^{-im\theta} \rangle \langle e^{+im\theta} A_2 e^{+in\theta} \rangle \right) \\ - \sum_{n=0}^{\infty} \sum_{m=1}^{\infty} \left( \langle e^{+in\theta} A_1 e^{+im\theta} \rangle \langle e^{-im\theta} A_2 e^{-in\theta} \rangle \right) \end{array} \right\} \tag{2.8}$$

Use (2.3) in the above this becomes,

$$F_2 = \frac{1}{4i} \left( \begin{array}{l} \sum_{n=1}^{\infty} \sum_{m=0}^{\infty} \big( \delta(-n+j-m) + \delta(-n-j-m) \big) \big( \delta(m+j+n) - \delta(m-j+n) \big) \\ - \sum_{n=0}^{\infty} \sum_{m=1}^{\infty} \big( \delta(n+j+m) + \delta(n-j+m) \big) \big( \delta(-m+j-n) - \delta(-m-j-n) \big) \end{array} \right) \tag{2.9}$$

where $\delta(k) = 1$ if $k = 0$ and $\delta(k) = 0$ if $k \neq 0$.

This yields,

$$F_2 = -\frac{1}{4i} \left[ \sum_{n=0}^{\infty} \sum_{m=1}^{\infty} \delta(m-j+n) + \sum_{n=1}^{\infty} \sum_{m=0}^{\infty} \delta(m-j+n) \right] \tag{2.10}$$

From this we obtain,

$$F_2 = -\frac{1}{4i} \left( \sum_{n=0}^{j-1} 1 + \sum_{n=1}^{j} 1 \right) = -\frac{j}{2i} \tag{2.11}$$

At this point we have an inconsistency. As discussed above if we use the cyclicity of the trace in (2.4) then $F_1 = 0$, but we have just shown that $F_2 \neq 0$ so that these results are in contradiction to (1.3).

To further understand this problem we will next examine the expression for $F_1$ that is given in Eq. (2.4) in more detail. Based on the previous discussion we obtain,

$$trace(a_1 a_2) = \frac{1}{(2\pi)^2} \sum_{n=0}^{\infty} \sum_{m=0}^{\infty} \left\langle e^{+in\theta} A_1 e^{-im\theta} \right\rangle \left\langle e^{+im\theta} A_2 e^{-in\theta} \right\rangle \tag{2.12}$$

Use (2.3) to obtain,

$$trace(a_1 a_2) = \frac{1}{4i} \sum_{n=0}^{\infty} \sum_{m=0}^{\infty} \begin{bmatrix} \delta(n+j-m) \\ +\delta(n-j-m) \end{bmatrix} \begin{bmatrix} \delta(m+j-n) \\ -\delta(m-j-n) \end{bmatrix} \tag{2.13}$$

$$trace(a_1 a_2) = \frac{1}{4i} \sum_{n=0}^{\infty} \sum_{m=0}^{\infty} \begin{bmatrix} \delta(n-j-m) \\ -\delta(n+j-m) \end{bmatrix} \tag{2.14}$$

Next switch the "dummy" indices $n$ and $m$ to obtain,

$$trace(a_1 a_2) = \frac{1}{4i} \sum_{n=0}^{\infty} \sum_{m=0}^{\infty} \begin{bmatrix} \delta(m-j-n) \\ -\delta(m+j-n) \end{bmatrix} = -\frac{1}{4i} \sum_{n=0}^{\infty} \sum_{m=0}^{\infty} \begin{bmatrix} \delta(n-j-m) \\ -\delta(n+j-m) \end{bmatrix} \tag{2.15}$$

Comparing the above two equations we have $trace(a_1 a_2) = -trace(a_1 a_2)$ which yields $trace(a_1 a_2) = 0$. Similarly it can be shown that $trace(a_2 a_1) = 0$. This confirms our original result that $F_1 = 0$.

## 3. Redefining the trace.

In this section we will show that we can eliminate this inconsistency by redefining the trace operation. We will redefine the trace operator as follows,

$$trace_L(A_{ij}) \underset{L \to \infty}{=} \sum_{n=-L}^{n=+L} A_{nn} \tag{3.1}$$

Note we have written it as $trace_L$ instead of $trace$ to distinguish it from the definition of the trace as given by (2.7). The key difference between the two definitions is that instead of taken the summation from $-\infty$ to $+\infty$ we take it from $-L$ to $+L$ and let $L \to \infty$ at the end of the calculation. In this case Eq. (2.12) is re-written as,

$$trace_L(a_1 a_2) \underset{L\to\infty}{=} \frac{1}{(2\pi)^2} \sum_{n=0}^{L} \sum_{m=0}^{\infty} \langle e^{+in\theta} A_1 e^{-im\theta} \rangle \langle e^{+im\theta} A_2 e^{-in\theta} \rangle \qquad (3.2)$$

which can be evaluated to obtain,

$$trace_L(a_1 a_2) \underset{L\to\infty}{=} \frac{1}{4i}\left(\sum_{n=0}^{L}\sum_{m=0}^{\infty} \delta(n-j-m) - \sum_{n=0}^{L}\sum_{m=0}^{\infty} \delta(n+j-m)\right) \qquad (3.3)$$

$$trace_L(a_1 a_2) \underset{L\to\infty}{=} \frac{1}{4i}\left(\sum_{n=j}^{L}(1) - \sum_{n=0}^{L}(1)\right) = \frac{1}{4i}\left((L+1-j)-(L+1)\right) = \frac{-j}{4i} \qquad (3.4)$$

Similarly,

$$trace_L(a_2 a_1) \underset{L\to\infty}{=} \frac{+j}{4i} \qquad (3.5)$$

In this case $F_1 = -j/2i$ and referring to (2.11) we obtain $F_1 = F_2$. Therefore the inconsistency is resolved by defining the trace operation according to (3.1) instead of (2.7).

Next examine $F_3$ which is given by Eq. (1.6). This relationship shows up in a number of papers [2,3,4]. Rearrange terms in (1.6) and use $J \cdot J = 1$ to obtain,

$$F_3 = \frac{1}{4} trace\begin{pmatrix}(A_1 J A_2 - J A_1 A_2) \\ -(A_1 A_2 J - J A_1 J A_2 J)\end{pmatrix} \qquad (3.6)$$

Next use (2.3) to obtain $A_1 A_2 = \sin(2j\theta)/2$. From this we have $\langle A_1 A_2 \rangle = 0$. From this it follows that $trace(J A_1 A_2) = trace(A_1 A_2 J) = 0$. Therefore (3.6) becomes,

$$F_3 = \frac{1}{2} trace(A_1 J A_2) \qquad (3.7)$$

where we have used $trace(J A_1 J A_2 J) = trace(A_1 J A_2)$. Proceeding as before we obtain

$$trace(A_1 J A_2) = \frac{1}{2i} \sum_{n=-\infty}^{n=+\infty} \sum_{m=0}^{+\infty} \left(\delta(m+j-n) - \delta(n+j-m)\right) \qquad (3.8)$$

Do the summation over the index $n$ first to obtain,

$$trace(A_1 J A_2) = \frac{1}{2i} \sum_{m=0}^{+\infty}(1-1) = 0 \qquad (3.9)$$

This gives us $F_3 = 0$. Since we have shown that $F_2 \neq 0$ and $F_1 \neq 0$ we have an inconsistency with Eq. (1.3). However if we define the trace according to (3.1) then (3.8) becomes,

$$trace_L\left(A_1 J A_2\right)\underset{L\to\infty}{=} \frac{1}{2i}\sum_{n=-L}^{n=+L}\sum_{m=0}^{+\infty}\left(\delta(m+j-n)-\delta(n+j-m)\right) \tag{3.10}$$

This can be evaluated to obtain,

$$trace_L\left(A_1 J A_2\right)\underset{L\to\infty}{=} -\frac{j}{i} \tag{3.11}$$

Use this in (3.7) to obtain $F_3 = -j/2i$ which is consistent with the previous result for $F_2$ and $F_1$.

Also note that if we had naively used the cyclicity of the trace in (3.6) we would have obtained,

$$F_3 = \frac{1}{2}trace\left(J[A_2, A_1]\right) = 0 \tag{3.12}$$

since $A_1$ and $A_2$ commute when defined by (2.3).

## 4. Conclusion.

We have examined an expression for the the Lie algebra cocycle corresponding to the central extension of the group $GL_{res}(H)$ as present in Section (6.6) of the book *Loop Groups* [1]. It has been shown that there will be ambiguity in calculating this quantity unless the trace operation is rigorously defined per Eq. (3.1).

## References.